\documentclass[%
 reprint,
 amsmath,amssymb,
 aps,
]{revtex4-1}

\usepackage{graphicx}
\usepackage{dcolumn}
\usepackage{bm}

\begin{document}

\preprint{APS/123-QED}

\title{Localization in covariance matrices of coupled heterogenous Ornstein-Uhlenbeck processes}

\author{Paolo Barucca}
 \email{baruccap@gmail.com}
\affiliation{%
 Dipartimento di Fisica, Universit\`a La Sapienza, P.le A. Moro 2, I-00185 Roma, Italy \\
}

\date{\today}

\begin{abstract}
We define a random-matrix ensemble given by the infinite-time covariance matrices of Ornstein-Uhlenbeck processes at different temperatures coupled by a Gaussian symmetric matrix. The spectral properties of this ensemble are shown to be in qualitative agreement with some stylized facts of financial markets. Through the presented model formulas are given for the analysis of heterogeneous time-series. Furthermore evidence for a localization transition in eigenvectors related to small and large eigenvalues in cross-correlations analysis of this model is found and a simple explanation of localization phenomena in financial time-series is provided. Finally we identify both in our model and in real financial data an inverted-bell effect in correlation between localized components and their local temperature: high and low temperature/volatility components are the most localized ones. 
\end{abstract}

\maketitle

\section{Introduction}
Complex systems are hard to analyse since by definition the interactions among their components are not easily connected with their behaviours \cite{pan2007collective}. 
In these systems the absence of a well-defined general model makes correlation analysis an irreplaceable, if not unique, compass \cite{podobnik2010time,biondo2013random}.
Furthermore in these systems the presence of noise makes benchmarking important and random matrix theory (RMT) is fundamental to check the statistical validity of pair-correlations. \\
RMT has mainly focused on the effects of the finite lengths of time series. In particular a careful analysis has been carried out on the spectral properties of random matrices in the case where the number of variables $N$ is large and the length of the signal $M$ is comparable, i.e. with a finite ratio $Q=M/N$  \cite{potters2005financial, plerou2002random, laloux1999noise, burda2004signal, utsugi2004random}. In this case the total time is not enough large for making the noise negligible: one needs to disentangle the properties induced by couplings from the ones brought by randomness.\\
Nevertheless time-series in complex systems are not only noisy and finite but also heterogeneous, which means their variances can be really different (i.e. the variance of one time series can be very different from the variance of another time series). More generally the marginal distribution of one variable may be qualitatively and quantitatively different from the one of another variable.\\
In finance, on which we will focus our considerations, the volatilities of different assets, i.e. the index of the percentage change in stock prices, have a very broad distribution \cite{cizeau1997volatility}, i.e. there is a strong heterogeneity between the returns of different assets. In recent studies it has been shown that this distribution is similar to a log-normal that is compatible with a fractal model of the market \cite{liu1999statistical, bouchaud2000apparent}. This feature has been included in models based on the random matrix Wishart ensemble to improve the comparison with real matrices \cite{burda2004free,burda2006spectral, akemann2010universal}. \\ 
Summarising complex systems are heterogeneous, disordered and noisy and they have a non-trivial relationship between interactions and correlations: carefully studied benchmarks are needed to gain a more detailed insight. 
In the following we will see how these different features are interconnected and we will point out how important is to consider them together in order to predict their effects on cross-correlation analysis. \\
The aim of this article is to observe the consequences of heterogeneity in a simple ad hoc model that allows to explicitly compute the relation between couplings and correlations. \\
In Section II we start from the basic dynamical model given by a set of independent Ornstein-Uhlenbeck (OU) processes at different temperatures. Then we turn to the interacting case where the OU processes are coupled through a given matrix. The ensemble we consider is the one given by the infinite-time covariance matrices of OU processes at different temperatures coupled by a Gaussian symmetric matrix. We also consider the stationary distribution of the time-series induced and show the relation with the known Wishart-Laguerre ensemble of random matrices. \\ 
In Section III we show the results of numerical simulations in the asymptotic limit. Varying heterogeneity we compute the spectral density of eigenvalues, the inverse participation ratio (IPR), a standard index of eigenvectors localization \cite{edwards1972numerical},  and the component participation ratio (CPR), that defines the contribution of a given component on all the eigenvectors. We check this ensemble properties both in averaged and single-sample eigenvectors. Moreover we identify a steep change in eigenvector localization driven by heterogeneity, that might be an indicator for a transition from an extended phase towards a localized phase in the eigenvectors of the cross-correlation matrix of the model. 
Finally we discuss the results both with respect to the known spectral properties of random matrix models and with the real localization properties widely observed in financial data \cite{chakraborti2011econophysics, plerou1999universal} and give theoretical perspectives.

\section{Coupled heterogeneous OU process}
\subsection{Indipendent OU processes}
In the following we will consider signals extracted from the equilibrium distribution of a continuous-time stochastic dynamics. The interest of this model for applications relies on the hypothesis that in complex systems observations are samplings from a complicated noisy dynamics as, for instance, in finance daily prices are the result of all the small price adjustments given by all the transactions. \\
We would like to stress though that we do not want to model a particular asset dynamics in detail: each class of assets may require a different dynamics and more complicated non-linear interaction terms that would not allow to give explicit formulas for the direct, from couplings to correlations, and inverse problem, from correlations to couplings. \\ 
The aim is to construct a null-model including a specific parametrization that separates couplings and temperatures in order to explicitly distinguish their role on the covariance matrix. 
We start our analysis from a limit-case, the noisy dynamics of $N$ independent variables $x = \{x_1, x_2 ... x_N\} $ following a standard OU process with a set of $N$ temperatures $T = \{T_1, T_2 ... T_N\} $ :
\begin{equation}
\dot{x_{i}} = -x_{i}+ \sqrt{T_{i}}\eta_{i}(t),
\end{equation}
where $\eta_{i}(t)$ is a delta-correlated Gaussian noise with $\langle\eta_{i}(t)\eta_{j}(t')\rangle=2\delta_{ij}\delta(t-t')$.
In this case the marginal equilibrium distribution for $x_i$ is $P_i(x_i)$:
\begin{equation}
P_i(x_{i}) = \frac{e^{-\frac{x_{i}^2}{2T_{i}} }}{\sqrt{2\pi T_{i}}}
\end{equation}
If we know sample the values of all the $x_i$'s at $M$ times we can compute the empirical covariance coefficients, $C_{ij} = \overline{x_{i} x_{j}} - \overline{x_i}\,\overline{x_j}$, where $\overline{\cdot} $ indicates the average over the $M$ sampled times. \\
For an infinite value of the ratio $Q=M/N$ the covariance matrix converges towards a diagonal one, $C_{ij} = T_i\delta_{ij}$. Meanwhile for a finite value of $Q$ the off-diagonal elements of $C$ are $N(N-1)/2$ Gaussian variables with zero mean and variance $\frac{T_iT_j}{M}$. \\
In this case the Pearson correlation-matrix $c_{ij} = C_{ij}/\sqrt{C_{ii}C_{jj}}$ has exactly the same statistics of a matrix extracted from the widely-used Wishart-Laguerre ensemble of random matrices since its elements are the pair-correlations of $N$ normally distributed signals of length $M$. 
The heterogeneity we have put in the dynamics plays no role in the correlation matrix in this case.

\subsection{Coupled OU processes}
The generalisation to the coupled case is interesting. The dynamics now verifies:
\begin{equation}\label{brownie}
\dot{x_{i}} = -\underset{j}{\sum}J_{ij} x_{j}+ \sqrt{T_{i}}\eta_{i}(t),
\end{equation}
where $J_{ij}$ is symmetric and positive-definite in order to ensure a finite limit to the process. In the following we will focus our analysis on the asymptotic limit since in the present work we are not interested in the consequences of the interplay of finite $Q$ and heterogeneity but solely on the consequences of the latter.
In this system there are two different methods \cite{risken1984fokker} to obtain a closed formula for the asymptotic covariance matrix, $C_{ij}=\langle x_{i}x_{j}\rangle$, as a function of couplings and temperatures ($\langle \cdot \rangle $ indicates the average over an infinite time). Starting from the dynamics with a few standard steps it is possible to find the implicit formula :
\begin{equation}\label{Cov}
\{C,J\}=2\tilde{T},
\end{equation}
where $\tilde{T}_{ij}=T_{i}\delta_{ij}$, and $\{\cdot,\cdot\}$ denotes
the matrix anti-commutator. 
From the spectral decomposition of $J$ it is possible to find a set of explicit formulas for the elements of $C_{ij}$:
\begin{equation}\label{Cov2}
C_{ij}=2\underset{a,b}{\sum}\frac{u_{i}^{a}u_{j}^{b}}{\lambda_{a}+\lambda_{b}}\underset{k}{\sum}u_{k}^{a}u_{k}^{b}T_{k},
\end{equation}
where $u_{i}^{a}$ is the $i$th component $a$th eigenvector of $J$
and $\lambda_{a}$ is the $a$-th eigenvalue.
In (\ref{Cov}) $C$ and $J$ appear in a symmetric form and the same symmetry must hold also in (\ref{Cov2}). This fact implies that (\ref{Cov2}) can be used to solve the inverse problem for this system, that is finding the couplings $J$ given the covariances $C$. This symmetry is not surprising since it holds also in the familiar homogeneous case where $C=J^{-1}$, an ostensibly symmetric formula. In Appendix A we examine the consequences on the Pearson correlation matrix $c$ in the case of small couplings. \\
We have thus defined two different random-matrix ensembles: one, that we will examine in the next Section, is the set of infinite-time covariance matrices that are defined by formula (\ref{Cov2}) for coupling matrices $J$ sampled from a given random-matrix ensemble (for instance the Gaussian ensemble) and for sets of temperatures $T$ sampled from a distribution chosen at will, the other (Appendix B) is the set of finite-time empirical covariance matrices between signals sampled from the stationary distribution of the OU dynamics for a given infinite-time $C$. 
\section{Sampling matrices}
\begin{figure}
\includegraphics[scale=0.2]{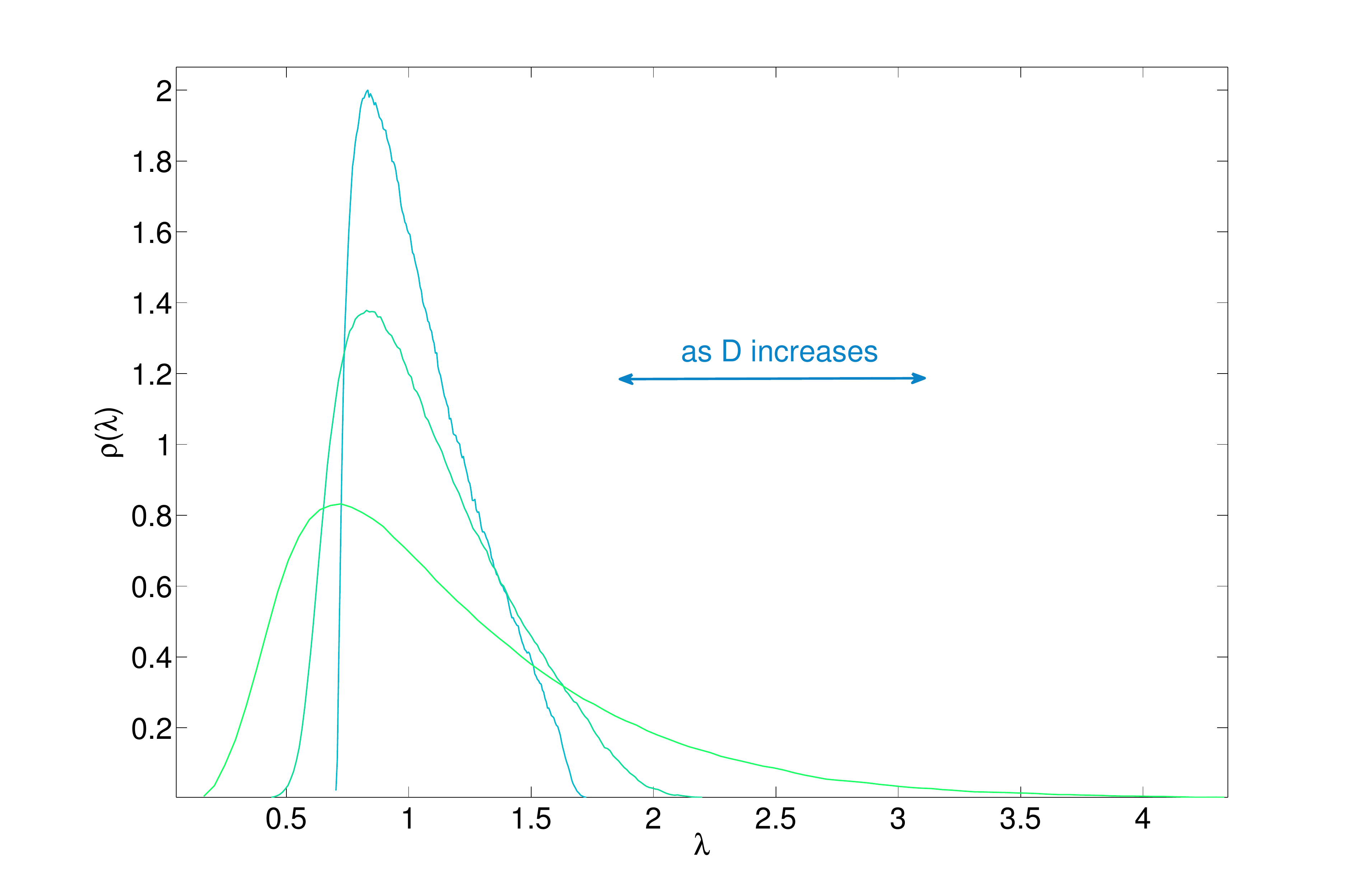}
\caption[spectrum]{For fixed $N=100$ and $\epsilon=0.2/\sqrt{N}$ we plot the spectral density of the correlation matrix $C$ for $D = 0.0, 0.2$ and $0.5$ obtained averaging over $10^3$ samples. Increasing $D$ we see that the lower edge of the spectrum becomes smaller and smaller and conversely that the higher edge increases.}
\label{Serie}
\end{figure}

Since we are interested in finding the consequences of heterogeneity we use straightly the infinite-time asymptotic formula (\ref{Cov2}) so that we avoid simulating the whole stochastic dynamics.
Thus we generate a random coupling matrix $J=I+\epsilon K$ where $I$
is the identity matrix, where $\epsilon$ is the strength of the coupling
among signals and $K$ and random Gaussian matrix whose elements
have variance $\frac{1}{N}$. $J$ must be positive-definite for any
$N$ so we eliminated samples with non-positive eigenvalues that have vanishing probability as $N$ goes to infinity. In principle it is possible to consider any kind of probability measure for couplings and temperatures, the main idea addressed here is to regard couplings as homogeneous so that temperatures are the only source of heterogeneity.
Since in the financial context temperatures represent volatilities that are typically log-normal distributed \cite{liu1999statistical, bouchaud2000apparent} we choose to draw them from this kind of distribution:
\begin{equation}
p(T)=\frac{1}{T}\frac{e^{-\frac{(\log T-\mu)^{2}}{2D^{2}}}}{\sqrt{2\pi D^{2}}}
\end{equation}
Namely we generate $N$ normally distributed random numbers, $\xi_{i}$,
and define $T_{i}=e^{\mu+D\xi_{i}}$. Then we fix $\epsilon$ and draw the coupling matrix $J$, diagonalise it and 
use (\ref{Cov2}) to obtain $C$. 
Varying $D$, $\epsilon$ and $N$ we observe some basic features of the $C$ matrix.
First we compute the eigenvalue distribution changing $N$ at fixed $D=1$ and $\epsilon=0.2/\sqrt{N}$ and we notice that, as $N$ increases, the distribution rapidly converges towards an infinite-size spectrum. 
Once this is verified we study the eigenvalue distribution varying $D$ alone. The spectrum spreads on both edges as is often observed in real data analysis Fig.\ref{Serie}. 
Thus introducing heterogeneity we have new eigenvalues, both small and large, so we enquiry the related eigenvectors and check whether they are statistically different from the ones in the homogeneous bulk of the spectrum.
\begin{figure}
\includegraphics[scale=0.2]{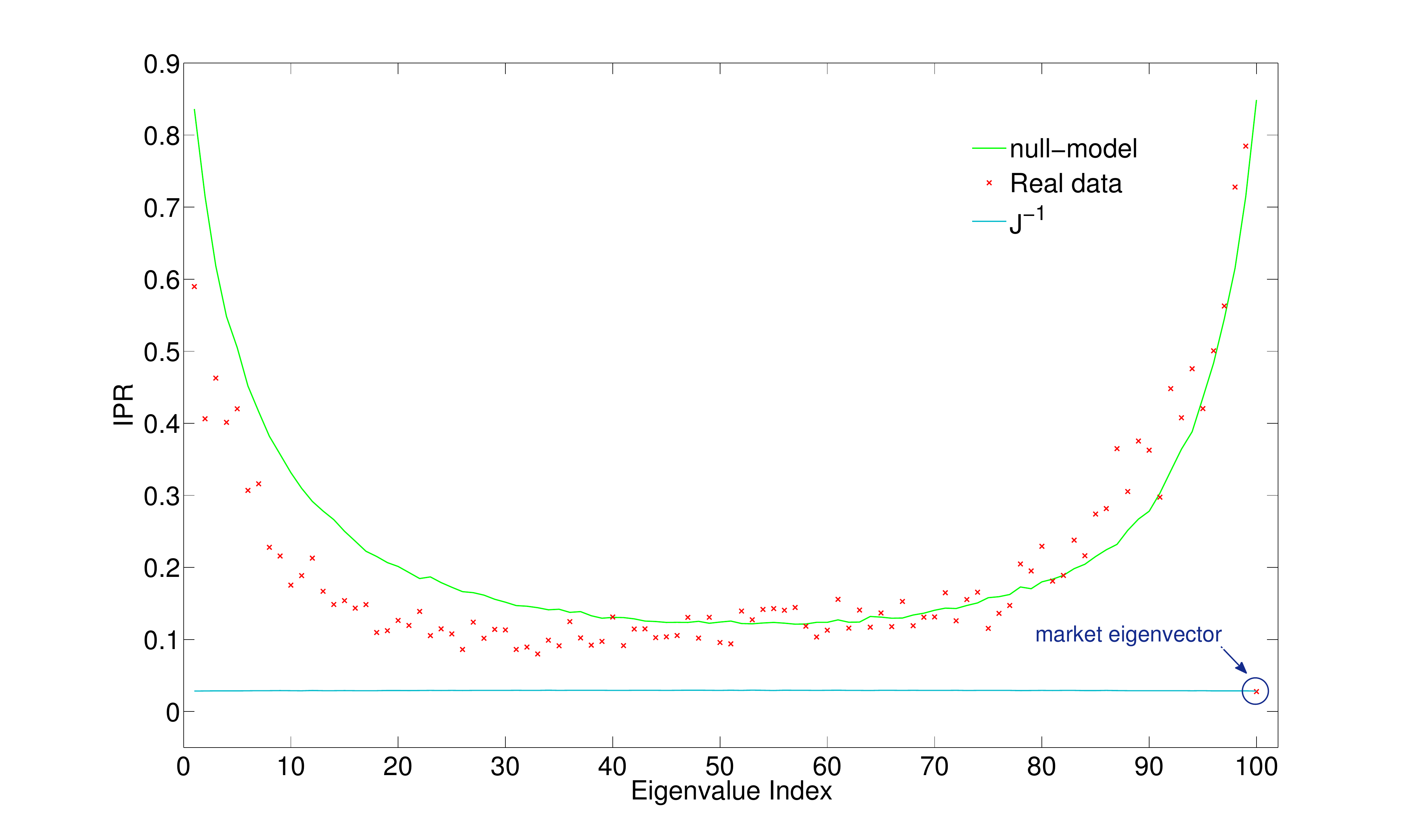}
\caption[avIPR]{For each ordered eigenvalue we plot the mean value of the n-th IPR versus the eigenvalue index averaged over $1000$ samples for a system size $N=100$ and a value of $\epsilon=0.2/\sqrt{N}$ for $D=.74$ and $\mu=7.74$ (as obtained from real data volatilities). Crosses show the IPR averaged over 10 matrices of daily asset returns from NYSE from the first of June $1987$ to the $31$ of December 1998. The $J^{-1}$ line is the equal temperatures case ($D=0$). We see that the largest eigenvector, representing the market, is extended and falls exactly on the $D=0$ line.}
\label{avIPR}
\end{figure}
We characterise the eigenvectors of $C$, $v_i^a$, through the IPR, a standard quantity in matrix analysis, defined by the formula:
\begin{equation}
IPR^a = \underset{i}{\sum}(v_i^a)^4
\end{equation}
Obviously $IPR$ values depend by the sample. Since we want to characterise its typical behaviour we take for each sample the set of ordered eigenvalues and consider their IPR, then the IPRs over samples Fig.\ref{avIPR}. Real data used are a set of $1017$ daily asset returns from NYSE from the first of June $1987$ to the $31$ of December 1998.
In order to compare qualitatively with data we fixed the values of the log-normal distribution by evaluating the mean and standard deviation of the  logarithm of returns variances, namely $\mu = \frac{1}{1017}\sum\limits_{k=1}^{1017}log(\sigma_{k})$ and $D = \frac{1}{1017}\sum\limits_{k=1}^{1017}(log(\sigma_{k})-\mu)^2$, where $\sigma_i$'s are the empirical variances.
The figure we obtain shows localization at the edges, a common feature observed in real data analysis. In particular the IPR shows agreement not only in the typical flat region related to the bulk where its value is fluctuating slightly over $3/N$ but also on the edges (see IPR in \cite{plerou1999universal}), where we observe the increasing of the $IPR$.

\begin{figure}
\includegraphics[scale=.2]{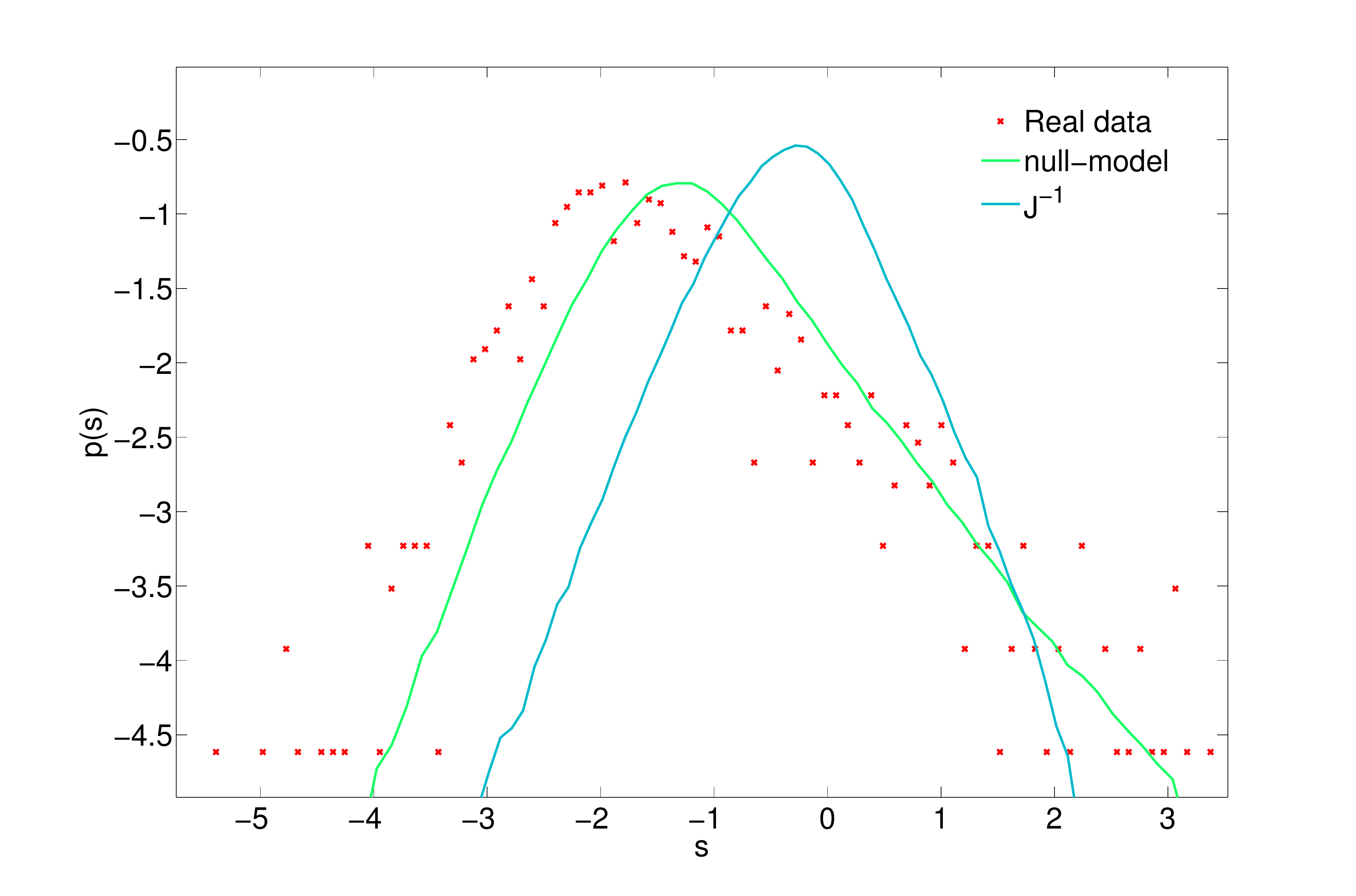}
\caption[trans]{Distribution of level spacings normalised by their mean value, $s_n = \frac{\lambda_{n+1} -\lambda_{n}}{\langle \lambda_{n+1} -\lambda_{n}\rangle} $, where $\lambda_n$ is the $n$-th eigenvalue of the covariance matrix. Data are presented in a log-log scale. Crosses show the IPR averaged over 10 matrices of daily asset returns from NYSE from the first of June $1987$ to the $31$ of December 1998. The $J^{-1}$ line is the equal temperatures case, $D=0$. Null-model data are averaged over $10^3$ samples for a system of size $N=100$ with $\mu$ and $D$ parameters obtained from real data. }
\label{levelSpace}
\end{figure}

We then evaluate level spacings, $\lambda_{n+1} -\lambda_{n}$,  where $\lambda_n$ is the $n$-th eigenvalue of the covariance matrix, and observe a clear left-shift in the spacings distribution \cite{shklovskii1993statistics,agam1995spectral}, mean that the skewness of spacings increases with heterogeneity Fig. \ref{levelSpace} approaching real data.

\begin{figure}
\includegraphics[scale=0.18]{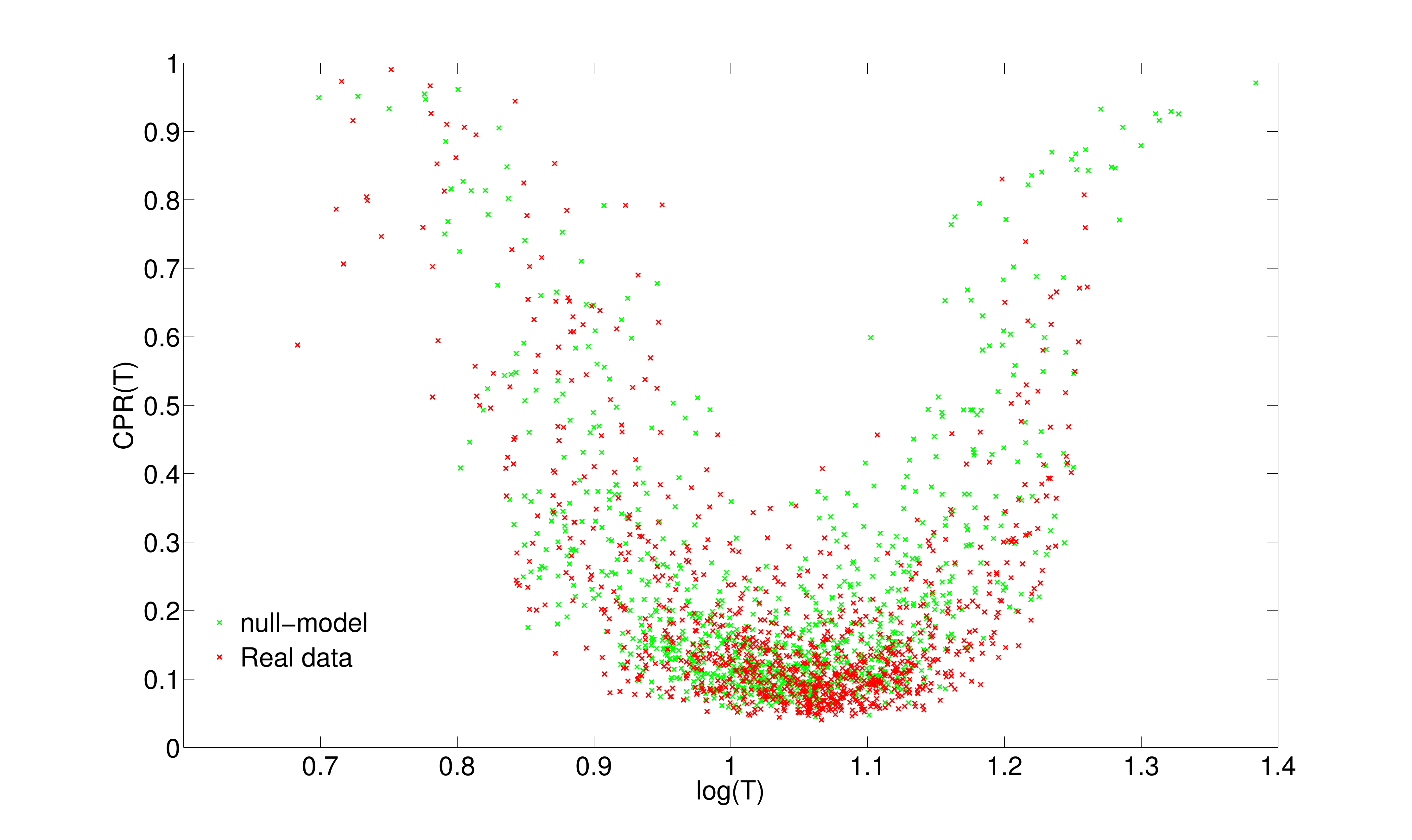}
\caption[trans]{Scatter plot of the components in the plane $(log(T_i), CPR_i)$. We can see an inverted-bell shape that is absent in GOE matrices, i.e. with no heterogeneity. Real and null-model data are over $10$ matrices of size $N=100$. For null-model data we used the values for $\mu$ and $D$ obtained from real data.}
\label{cprVsT}
\end{figure}

To observe the heterogeneity effect we also need to consider a matrix observable not depending on the eigenvector, such as IPR, but depending on the component so we study the component participation ratio that we define by the formula:
\begin{equation}
CPR_i = \underset{a}{\sum}(v_i^a)^4
\end{equation}
that is just the equivalent of the IPR for the change of basis matrix transposed. 
We investigate the relation between CPR and heterogeneity evaluating the correlations between CPR and both $T$ and $1/T$ by constructing the scatter plot $(\log(T_i), CPR_i)$. For real data we decided to approximate different temperatures with the diffusion terms \cite{siegert1998analysis} so we plot $(\log(D^{(e)}_{i}), CPR_i)$ Fig. \ref{cprVsT}, where $D_i^{(e)} =  \frac{1}{T}\sum\limits_{t=1}^{T-1}(r_i(t+1)-r_i(t))^2$ being $r_i(t)$ the return of asset $i$ at time $t$. The effect holds also considering variances versus CPR. \\
The inverted-bell shape indicates that high and low temperature/volatility components are the most localized ones. This result depends both on the presence of couplings and heterogenous temperatures/volatilities: with no couplings the covariance matrix would be diagonal and so all the eigenvectors would be localized and with too low heterogeneity the differences between diffusion terms would be negligible and would not affect localization so clearly. \\
An explanation for this effect can be achieved if we consider the uncoupled case where every eigenvector is sharply localized since the matrix is diagonal. If we now put a coupling between the components what happens is that the ones in the bulk with closer eigenvalues are likely to interact and spread while the ones on the edges are related to more isolated eigenvalues so are less likely to mix with others and will stay more localized. This picture should hold until the couplings are large enough to contrast the differences in temperature.  

\section{Conclusion}
We have analysed a simple model of complex systems
that provides a method for sampling random matrices. 
We have shown how our method gives results which are in agreement 
with eigenvector localization ubiquitous in real data. This model suggests that heterogeneity among signals is likely to cause localization, as indicated also by known random band models \cite{casati1990scaling, fyodorov1991scaling}. 
The analysis showed the peculiar characteristic that localization involves 
both the noisiest signals and the most deterministic ones, the inverted-bell effect.
Another interesting aspect is the heterogeneity effect in localization in the model proposed
showing a non-trivial transition from a coupling dominated phase, 
where spectral properties are the same as those of Wishart matrices, 
towards an heterogeneity dominated phase, where localization on the edges of the spectrum occurs.  A theoretical perspective is to establish whether the effect arises from a simple crossover or from a real phase transition, valid also in the thermodynamical limit, i.e. for infinite $N$, and possibly to characterise more in detail the two phases by examining also other matrix properties.
To improve the comparison with real data, especially in finance, another perspective is characterising the case of finite time-samplings, i.e. finite ratio $Q=M/N$ and check how the interplay of heterogeneities, couplings and finite time-samplings change the properties of the covariance matrix in a benchmark case. \\
The research leading to these results has received funding from the European Research Council under the European UnionÕs Seventh Framework Programme
(FP7/2007-2013) / ERC grant agreement No. 247328.
 \\

\section{Appendix A} 
We showed in the general case how couplings, covariances and heterogeneities are related. Here we show in a perturbative limit of small couplings what happens passing from the covariance to the correlation matrix. \\
We write $J = I + \epsilon K^{^{(1)}}$ where $I$ is the identity matrix, $K^{^{(1)}}$ is a random symmetric gaussian matrix and $\epsilon$ is an arbitrarily small real number. At first order in $\epsilon$ the covariance matrix must satisfy the perturbative expression $C = \tilde{T} + \epsilon \Sigma^{^{(1)}}$. Consequently $K^{^{(1)}}$ and $\Sigma^{^{(1)}}$ verify:
\begin{equation}\label{KS}
K^{^{(1)}}_{ij}(T_i + T_j)=-2\Sigma^{^{(1)}}_{ij}
\end{equation}
Furthermore $C_{ii} = T_i + \epsilon\Sigma^{^{(1)}}_{ii}$ so for the covariance matrix we have:
\begin{equation}
C_{ij}=T_i - \frac{\epsilon}{2}K^{^{(1)}}_{ij}(T_i+T_j)
\end{equation}
while the correlation matrix $c_{ij} = \frac{C_{ij}}{\sqrt{C_{ii} C_{jj}}}$ satisfies: 
\begin{equation}
c_{ij}=I - \frac{\epsilon}{2}\frac{K^{^{(1)}}_{ij}(T_i+T_j)}{\sqrt{T_i }\sqrt{T_j}}
\end{equation}
First-order expansion reveals a symmetry between $T$ and $1/T$ in the correlation matrix, that can be easily verified. This expansion allows us to consider a simplified random-matrix ensemble for the covariance matrices of weakly-coupled heterogeneous time-series for which analytical results can be obtained \cite{mythesis}. 
Moreover in case of strong heterogeneity, i.e. $T_i >> T_j$ $c_{ij}=c_{ji}=\frac{\epsilon}{2}\sqrt{\frac{T_i}{T_j}}K^{^{(1)}}_{ij}$, so if there is a low probability for a large value of $|K_{ij}^{^{(1)}}|$, the elements of the correlation matrix on the rows/columns related to variables with high or low temperature can be significantly bigger than the others. From the theory of Levy matrices \cite{cizeau1994theory} we know that large values of specific pair-correlation coefficients, i.e. a large $c_{mn}$, implies the presence of eigenvectors concentrated on the two components involved, e.g. $m$ and $n$. Moreover if the elements of a whole row are large compared to the rest of matrix there will be an eigenvector localized on the related component. 
An higher-order expansion shows the breaking of this high/low temperature symmetry in favour of the low-temperature components. 
At second order in $\epsilon$ we can write $C= T + \epsilon \Sigma^{^{(1)}}+\frac{\epsilon^2}{2} \Sigma^{^{(2)}}$ and $J= I + \epsilon K^{^{(1)}}+\frac{\epsilon^2}{2} K^{^{(2)}}$. This higher order expansion leads to the supplementary equation for $\Sigma^{^{(2)}}$:
\begin{equation}
2\Sigma^{^{(2)}}_{ij}= (T_i + T_j)(-K^{^{(2)}}_{ij} + \underset{k}{\sum}K^{^{(1)}}_{ik}K^{^{(1)}}_{kj}) + 2\underset{k}{\sum}K^{^{(1)}}_{ik}T_kK^{^{(1)}}_{kj},
\end{equation}
where we substituted $\Sigma^{^{(1)}}$ with the expression found at first-order (\ref{KS}).
If we know divide by $\sqrt{T_i}\sqrt{T_j}$ we obtain the second order correction to the correlation matrix $c$ that reads:
\begin{equation}
\frac{(T_i + T_j)}{\sqrt{T_i}\sqrt{T_j}}(-K^{^{(2)}}_{ij} + \underset{k}{\sum}K^{^{(1)}}_{ik}K^{^{(1)}}_{kj})+ \frac{2\underset{k}{\sum}K^{^{(1)}}_{ik}T_kK^{^{(1)}}_{kj}}{\sqrt{T_i}\sqrt{T_j}}.
\end{equation}
The first two terms remain unchanged if we substitute $T_i$ with $1/T_i$ but the third one does not, it breaks the symmetry in favour of elements related to components with low temperatures. 
We stress the fact that $\epsilon$ is small regardless the value of the system size $N$. If one performed the expansion for large $N$, then terms at all orders would have to be considered since at higher orders matrix multiplication would involve sums on an increasing number of elements.

\section{Appendix B}
For a given coupling matrix $J$ and set of temperatures $T$ the equilibrium distribution of the signals $x_i$ is a multivariate Gaussian, namely:
\begin{equation}
P(\{x_{i}\}| J, T) = \frac{\exp{{(-\frac{x^TC^{-1}x}{2}) }}}{\sqrt{(2\pi)^N \det{C} }}
\end{equation}
where $C$ is the covariance matrix, solution of eq. (\ref{Cov2}).
The empirical covariance matrix between signals extracted from this distribution defines a correlated Wishart ensemble \cite{marvcenko1967distribution, burda2005spectral, burda2006spectral,simon2004eigenvalue} whose peculiarity is the separation of the quenched disorders given by couplings and temperatures.

\section{Acknowledgements}
I want to thank C. Cammarota, B. Cerruti, A. Decelle, C. Lucibello, G.Parisi, J. Rocchi and B. Seoane for interesting discussions.

\bibliography{iprBib}{}

\begin{thebibliography}{28}%
\makeatletter
\providecommand \@ifxundefined [1]{%
 \@ifx{#1\undefined}
}%
\providecommand \@ifnum [1]{%
 \ifnum #1\expandafter \@firstoftwo
 \else \expandafter \@secondoftwo
 \fi
}%
\providecommand \@ifx [1]{%
 \ifx #1\expandafter \@firstoftwo
 \else \expandafter \@secondoftwo
 \fi
}%
\providecommand \natexlab [1]{#1}%
\providecommand \enquote  [1]{``#1''}%
\providecommand \bibnamefont  [1]{#1}%
\providecommand \bibfnamefont [1]{#1}%
\providecommand \citenamefont [1]{#1}%
\providecommand \href@noop [0]{\@secondoftwo}%
\providecommand \href [0]{\begingroup \@sanitize@url \@href}%
\providecommand \@href[1]{\@@startlink{#1}\@@href}%
\providecommand \@@href[1]{\endgroup#1\@@endlink}%
\providecommand \@sanitize@url [0]{\catcode `\\12\catcode `\$12\catcode
  `\&12\catcode `\#12\catcode `\^12\catcode `\_12\catcode `\%12\relax}%
\providecommand \@@startlink[1]{}%
\providecommand \@@endlink[0]{}%
\providecommand \url  [0]{\begingroup\@sanitize@url \@url }%
\providecommand \@url [1]{\endgroup\@href {#1}{\urlprefix }}%
\providecommand \urlprefix  [0]{URL }%
\providecommand \Eprint [0]{\href }%
\providecommand \doibase [0]{http://dx.doi.org/}%
\providecommand \selectlanguage [0]{\@gobble}%
\providecommand \bibinfo  [0]{\@secondoftwo}%
\providecommand \bibfield  [0]{\@secondoftwo}%
\providecommand \translation [1]{[#1]}%
\providecommand \BibitemOpen [0]{}%
\providecommand \bibitemStop [0]{}%
\providecommand \bibitemNoStop [0]{.\EOS\space}%
\providecommand \EOS [0]{\spacefactor3000\relax}%
\providecommand \BibitemShut  [1]{\csname bibitem#1\endcsname}%
\let\auto@bib@innerbib\@empty
\bibitem [{\citenamefont {Pan}\ and\ \citenamefont
  {Sinha}(2007)}]{pan2007collective}%
  \BibitemOpen
  \bibfield  {author} {\bibinfo {author} {\bibfnamefont {R.~K.}\ \bibnamefont
  {Pan}}\ and\ \bibinfo {author} {\bibfnamefont {S.}~\bibnamefont {Sinha}},\
  }\href@noop {} {\bibfield  {journal} {\bibinfo  {journal} {Phys. Rev. E}\
  }\textbf {\bibinfo {volume} {76}},\ \bibinfo {pages} {046116} (\bibinfo
  {year} {2007})}\BibitemShut {NoStop}%
\bibitem [{\citenamefont {Podobnik}\ \emph {et~al.}(2010)\citenamefont
  {Podobnik}, \citenamefont {Wang}, \citenamefont {Horvatic}, \citenamefont
  {Grosse},\ and\ \citenamefont {Stanley}}]{podobnik2010time}%
  \BibitemOpen
  \bibfield  {author} {\bibinfo {author} {\bibfnamefont {B.}~\bibnamefont
  {Podobnik}}, \bibinfo {author} {\bibfnamefont {D.}~\bibnamefont {Wang}},
  \bibinfo {author} {\bibfnamefont {D.}~\bibnamefont {Horvatic}}, \bibinfo
  {author} {\bibfnamefont {I.}~\bibnamefont {Grosse}}, \ and\ \bibinfo {author}
  {\bibfnamefont {H.~E.}\ \bibnamefont {Stanley}},\ }\href@noop {} {\bibfield
  {journal} {\bibinfo  {journal} {EPL (Europhysics Letters)}\ }\textbf
  {\bibinfo {volume} {90}},\ \bibinfo {pages} {68001} (\bibinfo {year}
  {2010})}\BibitemShut {NoStop}%
\bibitem [{\citenamefont {Biondo}\ \emph {et~al.}(2013)\citenamefont {Biondo},
  \citenamefont {Pluchino}, \citenamefont {Rapisarda},\ and\ \citenamefont
  {Helbing}}]{biondo2013random}%
  \BibitemOpen
  \bibfield  {author} {\bibinfo {author} {\bibfnamefont {A.~E.}\ \bibnamefont
  {Biondo}}, \bibinfo {author} {\bibfnamefont {A.}~\bibnamefont {Pluchino}},
  \bibinfo {author} {\bibfnamefont {A.}~\bibnamefont {Rapisarda}}, \ and\
  \bibinfo {author} {\bibfnamefont {D.}~\bibnamefont {Helbing}},\ }\href@noop
  {} {\bibfield  {journal} {\bibinfo  {journal} {PloS one}\ }\textbf {\bibinfo
  {volume} {8}},\ \bibinfo {pages} {e68344} (\bibinfo {year}
  {2013})}\BibitemShut {NoStop}%
\bibitem [{\citenamefont {Potters}\ \emph {et~al.}(2005)\citenamefont
  {Potters}, \citenamefont {Bouchaud},\ and\ \citenamefont
  {Laloux}}]{potters2005financial}%
  \BibitemOpen
  \bibfield  {author} {\bibinfo {author} {\bibfnamefont {M.}~\bibnamefont
  {Potters}}, \bibinfo {author} {\bibfnamefont {J.-P.}\ \bibnamefont
  {Bouchaud}}, \ and\ \bibinfo {author} {\bibfnamefont {L.}~\bibnamefont
  {Laloux}},\ }\href@noop {} {\bibfield  {journal} {\bibinfo  {journal} {arXiv
  physics/0507111, Financial applications of random matrix theory: Old laces
  and new pieces}\ } (\bibinfo {year} {2005})}\BibitemShut {NoStop}%
\bibitem [{\citenamefont {Plerou}\ \emph {et~al.}(2002)\citenamefont {Plerou},
  \citenamefont {Gopikrishnan}, \citenamefont {Rosenow}, \citenamefont
  {Amaral}, \citenamefont {Guhr},\ and\ \citenamefont
  {Stanley}}]{plerou2002random}%
  \BibitemOpen
  \bibfield  {author} {\bibinfo {author} {\bibfnamefont {V.}~\bibnamefont
  {Plerou}}, \bibinfo {author} {\bibfnamefont {P.}~\bibnamefont
  {Gopikrishnan}}, \bibinfo {author} {\bibfnamefont {B.}~\bibnamefont
  {Rosenow}}, \bibinfo {author} {\bibfnamefont {L.~A.~N.}\ \bibnamefont
  {Amaral}}, \bibinfo {author} {\bibfnamefont {T.}~\bibnamefont {Guhr}}, \ and\
  \bibinfo {author} {\bibfnamefont {H.~E.}\ \bibnamefont {Stanley}},\
  }\href@noop {} {\bibfield  {journal} {\bibinfo  {journal} {Phys. Rev. E}\
  }\textbf {\bibinfo {volume} {65}},\ \bibinfo {pages} {066126} (\bibinfo
  {year} {2002})}\BibitemShut {NoStop}%
\bibitem [{\citenamefont {Laloux}\ \emph {et~al.}(1999)\citenamefont {Laloux},
  \citenamefont {Cizeau}, \citenamefont {Bouchaud},\ and\ \citenamefont
  {Potters}}]{laloux1999noise}%
  \BibitemOpen
  \bibfield  {author} {\bibinfo {author} {\bibfnamefont {L.}~\bibnamefont
  {Laloux}}, \bibinfo {author} {\bibfnamefont {P.}~\bibnamefont {Cizeau}},
  \bibinfo {author} {\bibfnamefont {J.-P.}\ \bibnamefont {Bouchaud}}, \ and\
  \bibinfo {author} {\bibfnamefont {M.}~\bibnamefont {Potters}},\ }\href@noop
  {} {\bibfield  {journal} {\bibinfo  {journal} {Phys. Rev. Lett.}\ }\textbf
  {\bibinfo {volume} {83}},\ \bibinfo {pages} {1467} (\bibinfo {year}
  {1999})}\BibitemShut {NoStop}%
\bibitem [{\citenamefont {Burda}\ and\ \citenamefont
  {Jurkiewicz}(2004)}]{burda2004signal}%
  \BibitemOpen
  \bibfield  {author} {\bibinfo {author} {\bibfnamefont {Z.}~\bibnamefont
  {Burda}}\ and\ \bibinfo {author} {\bibfnamefont {J.}~\bibnamefont
  {Jurkiewicz}},\ }\href@noop {} {\bibfield  {journal} {\bibinfo  {journal}
  {Physica A: Stat. Mech.}\ }\textbf {\bibinfo {volume} {344}},\ \bibinfo
  {pages} {67} (\bibinfo {year} {2004})}\BibitemShut {NoStop}%
\bibitem [{\citenamefont {Utsugi}\ \emph {et~al.}(2004)\citenamefont {Utsugi},
  \citenamefont {Ino},\ and\ \citenamefont {Oshikawa}}]{utsugi2004random}%
  \BibitemOpen
  \bibfield  {author} {\bibinfo {author} {\bibfnamefont {A.}~\bibnamefont
  {Utsugi}}, \bibinfo {author} {\bibfnamefont {K.}~\bibnamefont {Ino}}, \ and\
  \bibinfo {author} {\bibfnamefont {M.}~\bibnamefont {Oshikawa}},\ }\href@noop
  {} {\bibfield  {journal} {\bibinfo  {journal} {Phys. Rev. E}\ }\textbf
  {\bibinfo {volume} {70}},\ \bibinfo {pages} {026110} (\bibinfo {year}
  {2004})}\BibitemShut {NoStop}%
\bibitem [{\citenamefont {Cizeau}\ \emph {et~al.}(1997)\citenamefont {Cizeau},
  \citenamefont {Liu}, \citenamefont {Meyer}, \citenamefont {Peng},\ and\
  \citenamefont {Eugene~Stanley}}]{cizeau1997volatility}%
  \BibitemOpen
  \bibfield  {author} {\bibinfo {author} {\bibfnamefont {P.}~\bibnamefont
  {Cizeau}}, \bibinfo {author} {\bibfnamefont {Y.}~\bibnamefont {Liu}},
  \bibinfo {author} {\bibfnamefont {M.}~\bibnamefont {Meyer}}, \bibinfo
  {author} {\bibfnamefont {C.-K.}\ \bibnamefont {Peng}}, \ and\ \bibinfo
  {author} {\bibfnamefont {H.}~\bibnamefont {Eugene~Stanley}},\ }\href@noop {}
  {\bibfield  {journal} {\bibinfo  {journal} {Physica A: Stat. Mech.}\ }\textbf
  {\bibinfo {volume} {245}},\ \bibinfo {pages} {441} (\bibinfo {year}
  {1997})}\BibitemShut {NoStop}%
\bibitem [{\citenamefont {Liu}\ \emph {et~al.}(1999)\citenamefont {Liu},
  \citenamefont {Gopikrishnan}, \citenamefont {Cizeau},\ and\ \citenamefont
  {Stanley}}]{liu1999statistical}%
  \BibitemOpen
  \bibfield  {author} {\bibinfo {author} {\bibfnamefont {Y.}~\bibnamefont
  {Liu}}, \bibinfo {author} {\bibfnamefont {P.}~\bibnamefont {Gopikrishnan}},
  \bibinfo {author} {\bibfnamefont {P.}~\bibnamefont {Cizeau}, \bibfnamefont
  {Meyer}}, \ and\ \bibinfo {author} {\bibfnamefont {H.~E.}\ \bibnamefont
  {Stanley}},\ }\href@noop {} {\bibfield  {journal} {\bibinfo  {journal} {Phys.
  Rev. E}\ }\textbf {\bibinfo {volume} {60}},\ \bibinfo {pages} {1390}
  (\bibinfo {year} {1999})}\BibitemShut {NoStop}%
\bibitem [{\citenamefont {Bouchaud}\ \emph {et~al.}(2000)\citenamefont
  {Bouchaud}, \citenamefont {Potters},\ and\ \citenamefont
  {Meyer}}]{bouchaud2000apparent}%
  \BibitemOpen
  \bibfield  {author} {\bibinfo {author} {\bibfnamefont {J.-P.}\ \bibnamefont
  {Bouchaud}}, \bibinfo {author} {\bibfnamefont {M.}~\bibnamefont {Potters}}, \
  and\ \bibinfo {author} {\bibfnamefont {M.}~\bibnamefont {Meyer}},\
  }\href@noop {} {\bibfield  {journal} {\bibinfo  {journal} {The European
  Physical J. B (Cond. Matt.)}\ }\textbf {\bibinfo {volume} {13}},\ \bibinfo
  {pages} {595} (\bibinfo {year} {2000})}\BibitemShut {NoStop}%
\bibitem [{\citenamefont {Burda}\ \emph {et~al.}(2004)\citenamefont {Burda},
  \citenamefont {Jurkiewicz}, \citenamefont {Nowak}, \citenamefont {Papp},\
  and\ \citenamefont {Zahed}}]{burda2004free}%
  \BibitemOpen
  \bibfield  {author} {\bibinfo {author} {\bibfnamefont {Z.}~\bibnamefont
  {Burda}}, \bibinfo {author} {\bibfnamefont {J.}~\bibnamefont {Jurkiewicz}},
  \bibinfo {author} {\bibfnamefont {M.~A.}\ \bibnamefont {Nowak}}, \bibinfo
  {author} {\bibfnamefont {G.}~\bibnamefont {Papp}}, \ and\ \bibinfo {author}
  {\bibfnamefont {I.}~\bibnamefont {Zahed}},\ }\href@noop {} {\bibfield
  {journal} {\bibinfo  {journal} {Physica A: Stat. Mech.}\ }\textbf {\bibinfo
  {volume} {343}},\ \bibinfo {pages} {694} (\bibinfo {year}
  {2004})}\BibitemShut {NoStop}%
\bibitem [{\citenamefont {Burda}\ \emph {et~al.}(2006)\citenamefont {Burda},
  \citenamefont {G{\"o}rlich},\ and\ \citenamefont
  {Wac{\l}aw}}]{burda2006spectral}%
  \BibitemOpen
  \bibfield  {author} {\bibinfo {author} {\bibfnamefont {Z.}~\bibnamefont
  {Burda}}, \bibinfo {author} {\bibfnamefont {A.~T.}\ \bibnamefont
  {G{\"o}rlich}}, \ and\ \bibinfo {author} {\bibfnamefont {B.}~\bibnamefont
  {Wac{\l}aw}},\ }\href@noop {} {\bibfield  {journal} {\bibinfo  {journal}
  {Phys. Rev. E}\ }\textbf {\bibinfo {volume} {74}},\ \bibinfo {pages} {041129}
  (\bibinfo {year} {2006})}\BibitemShut {NoStop}%
\bibitem [{\citenamefont {Akemann}\ \emph {et~al.}(2010)\citenamefont
  {Akemann}, \citenamefont {Fischmann},\ and\ \citenamefont
  {Vivo}}]{akemann2010universal}%
  \BibitemOpen
  \bibfield  {author} {\bibinfo {author} {\bibfnamefont {G.}~\bibnamefont
  {Akemann}}, \bibinfo {author} {\bibfnamefont {J.}~\bibnamefont {Fischmann}},
  \ and\ \bibinfo {author} {\bibfnamefont {P.}~\bibnamefont {Vivo}},\
  }\href@noop {} {\bibfield  {journal} {\bibinfo  {journal} {Physica A: Stat.
  Mech.}\ }\textbf {\bibinfo {volume} {389}},\ \bibinfo {pages} {2566}
  (\bibinfo {year} {2010})}\BibitemShut {NoStop}%
\bibitem [{\citenamefont {Edwards}\ and\ \citenamefont
  {Thouless}(1972)}]{edwards1972numerical}%
  \BibitemOpen
  \bibfield  {author} {\bibinfo {author} {\bibfnamefont {J.}~\bibnamefont
  {Edwards}}\ and\ \bibinfo {author} {\bibfnamefont {D.}~\bibnamefont
  {Thouless}},\ }\href@noop {} {\bibfield  {journal} {\bibinfo  {journal}
  {Journal of Physics C: Solid State Physics}\ }\textbf {\bibinfo {volume}
  {5}},\ \bibinfo {pages} {807} (\bibinfo {year} {1972})}\BibitemShut {NoStop}%
\bibitem [{\citenamefont {Chakraborti}\ \emph {et~al.}(2011)\citenamefont
  {Chakraborti}, \citenamefont {Toke}, \citenamefont {Patriarca},\ and\
  \citenamefont {Abergel}}]{chakraborti2011econophysics}%
  \BibitemOpen
  \bibfield  {author} {\bibinfo {author} {\bibfnamefont {A.}~\bibnamefont
  {Chakraborti}}, \bibinfo {author} {\bibfnamefont {I.~M.}\ \bibnamefont
  {Toke}}, \bibinfo {author} {\bibfnamefont {M.}~\bibnamefont {Patriarca}}, \
  and\ \bibinfo {author} {\bibfnamefont {F.}~\bibnamefont {Abergel}},\
  }\href@noop {} {\bibfield  {journal} {\bibinfo  {journal} {Quant. Fin.}\
  }\textbf {\bibinfo {volume} {11}},\ \bibinfo {pages} {991} (\bibinfo {year}
  {2011})}\BibitemShut {NoStop}%
\bibitem [{\citenamefont {Plerou}\ \emph {et~al.}(1999)\citenamefont {Plerou},
  \citenamefont {Gopikrishnan}, \citenamefont {Rosenow}, \citenamefont
  {Amaral},\ and\ \citenamefont {Stanley}}]{plerou1999universal}%
  \BibitemOpen
  \bibfield  {author} {\bibinfo {author} {\bibfnamefont {V.}~\bibnamefont
  {Plerou}}, \bibinfo {author} {\bibfnamefont {P.}~\bibnamefont
  {Gopikrishnan}}, \bibinfo {author} {\bibfnamefont {B.}~\bibnamefont
  {Rosenow}}, \bibinfo {author} {\bibfnamefont {L.~A.~N.}\ \bibnamefont
  {Amaral}}, \ and\ \bibinfo {author} {\bibfnamefont {H.~E.}\ \bibnamefont
  {Stanley}},\ }\href@noop {} {\bibfield  {journal} {\bibinfo  {journal} {Phys.
  Rev. Lett.}\ }\textbf {\bibinfo {volume} {83}},\ \bibinfo {pages} {1471}
  (\bibinfo {year} {1999})}\BibitemShut {NoStop}%
\bibitem [{\citenamefont {Risken}(1984)}]{risken1984fokker}%
  \BibitemOpen
  \bibfield  {author} {\bibinfo {author} {\bibfnamefont {H.}~\bibnamefont
  {Risken}},\ }\href@noop {} {\emph {\bibinfo {title} {Fokker-Planck
  Equation}}}\ (\bibinfo  {publisher} {Springer},\ \bibinfo {year}
  {1984})\BibitemShut {NoStop}%
\bibitem [{\citenamefont {Shklovskii}\ \emph {et~al.}(1993)\citenamefont
  {Shklovskii}, \citenamefont {Shapiro}, \citenamefont {Sears}, \citenamefont
  {Lambrianides},\ and\ \citenamefont {Shore}}]{shklovskii1993statistics}%
  \BibitemOpen
  \bibfield  {author} {\bibinfo {author} {\bibfnamefont {B.}~\bibnamefont
  {Shklovskii}}, \bibinfo {author} {\bibfnamefont {B.}~\bibnamefont {Shapiro}},
  \bibinfo {author} {\bibfnamefont {B.}~\bibnamefont {Sears}}, \bibinfo
  {author} {\bibfnamefont {P.}~\bibnamefont {Lambrianides}}, \ and\ \bibinfo
  {author} {\bibfnamefont {H.}~\bibnamefont {Shore}},\ }\href@noop {}
  {\bibfield  {journal} {\bibinfo  {journal} {Physical Review B}\ }\textbf
  {\bibinfo {volume} {47}},\ \bibinfo {pages} {11487} (\bibinfo {year}
  {1993})}\BibitemShut {NoStop}%
\bibitem [{\citenamefont {Agam}\ \emph {et~al.}(1995)\citenamefont {Agam},
  \citenamefont {Altshuler},\ and\ \citenamefont {Andreev}}]{agam1995spectral}%
  \BibitemOpen
  \bibfield  {author} {\bibinfo {author} {\bibfnamefont {O.}~\bibnamefont
  {Agam}}, \bibinfo {author} {\bibfnamefont {B.~L.}\ \bibnamefont {Altshuler}},
  \ and\ \bibinfo {author} {\bibfnamefont {A.~V.}\ \bibnamefont {Andreev}},\
  }\href@noop {} {\bibfield  {journal} {\bibinfo  {journal} {Physical review
  letters}\ }\textbf {\bibinfo {volume} {75}},\ \bibinfo {pages} {4389}
  (\bibinfo {year} {1995})}\BibitemShut {NoStop}%
\bibitem [{\citenamefont {Siegert}\ \emph {et~al.}(1998)\citenamefont
  {Siegert}, \citenamefont {Friedrich},\ and\ \citenamefont
  {Peinke}}]{siegert1998analysis}%
  \BibitemOpen
  \bibfield  {author} {\bibinfo {author} {\bibfnamefont {S.}~\bibnamefont
  {Siegert}}, \bibinfo {author} {\bibfnamefont {R.}~\bibnamefont {Friedrich}},
  \ and\ \bibinfo {author} {\bibfnamefont {J.}~\bibnamefont {Peinke}},\
  }\href@noop {} {\bibfield  {journal} {\bibinfo  {journal} {Physics Letters
  A}\ }\textbf {\bibinfo {volume} {243}},\ \bibinfo {pages} {275} (\bibinfo
  {year} {1998})}\BibitemShut {NoStop}%
\bibitem [{\citenamefont {Casati}\ \emph {et~al.}(1990)\citenamefont {Casati},
  \citenamefont {Molinari},\ and\ \citenamefont
  {Izrailev}}]{casati1990scaling}%
  \BibitemOpen
  \bibfield  {author} {\bibinfo {author} {\bibfnamefont {G.}~\bibnamefont
  {Casati}}, \bibinfo {author} {\bibfnamefont {L.}~\bibnamefont {Molinari}}, \
  and\ \bibinfo {author} {\bibfnamefont {F.}~\bibnamefont {Izrailev}},\
  }\href@noop {} {\bibfield  {journal} {\bibinfo  {journal} {Phys. Rev. Lett.}\
  }\textbf {\bibinfo {volume} {64}},\ \bibinfo {pages} {1851} (\bibinfo {year}
  {1990})}\BibitemShut {NoStop}%
\bibitem [{\citenamefont {Fyodorov}\ and\ \citenamefont
  {Mirlin}(1991)}]{fyodorov1991scaling}%
  \BibitemOpen
  \bibfield  {author} {\bibinfo {author} {\bibfnamefont {Y.~V.}\ \bibnamefont
  {Fyodorov}}\ and\ \bibinfo {author} {\bibfnamefont {A.~D.}\ \bibnamefont
  {Mirlin}},\ }\href@noop {} {\bibfield  {journal} {\bibinfo  {journal} {Phys.
  Rev. Lett.}\ }\textbf {\bibinfo {volume} {67}},\ \bibinfo {pages} {2405}
  (\bibinfo {year} {1991})}\BibitemShut {NoStop}%
\bibitem [{\citenamefont {Barucca}(2014)}]{mythesis}%
  \BibitemOpen
  \bibfield  {author} {\bibinfo {author} {\bibfnamefont {P.}~\bibnamefont
  {Barucca}},\ }\emph {\bibinfo {title} {Quenched heterogeneities in disordered
  systems}},\ \href@noop {} {Ph.D. thesis},\ \bibinfo  {school} {Sapienza
  University} (\bibinfo {year} {2014})\BibitemShut {NoStop}%
\bibitem [{\citenamefont {Cizeau}\ and\ \citenamefont
  {Bouchaud}(1994)}]{cizeau1994theory}%
  \BibitemOpen
  \bibfield  {author} {\bibinfo {author} {\bibfnamefont {P.}~\bibnamefont
  {Cizeau}}\ and\ \bibinfo {author} {\bibfnamefont {J.-P.}\ \bibnamefont
  {Bouchaud}},\ }\href@noop {} {\bibfield  {journal} {\bibinfo  {journal}
  {Phys. Rev. E}\ }\textbf {\bibinfo {volume} {50}},\ \bibinfo {pages} {1810}
  (\bibinfo {year} {1994})}\BibitemShut {NoStop}%
\bibitem [{\citenamefont {Mar{\v{c}}enko}\ and\ \citenamefont
  {Pastur}(1967)}]{marvcenko1967distribution}%
  \BibitemOpen
  \bibfield  {author} {\bibinfo {author} {\bibfnamefont {V.~A.}\ \bibnamefont
  {Mar{\v{c}}enko}}\ and\ \bibinfo {author} {\bibfnamefont {L.~A.}\
  \bibnamefont {Pastur}},\ }\href@noop {} {\bibfield  {journal} {\bibinfo
  {journal} {Sbornik: Mathematics}\ }\textbf {\bibinfo {volume} {1}},\ \bibinfo
  {pages} {457} (\bibinfo {year} {1967})}\BibitemShut {NoStop}%
\bibitem [{\citenamefont {Burda}\ \emph {et~al.}(2005)\citenamefont {Burda},
  \citenamefont {Jurkiewicz},\ and\ \citenamefont
  {Wac{\l}aw}}]{burda2005spectral}%
  \BibitemOpen
  \bibfield  {author} {\bibinfo {author} {\bibfnamefont {Z.}~\bibnamefont
  {Burda}}, \bibinfo {author} {\bibfnamefont {J.}~\bibnamefont {Jurkiewicz}}, \
  and\ \bibinfo {author} {\bibfnamefont {B.}~\bibnamefont {Wac{\l}aw}},\
  }\href@noop {} {\bibfield  {journal} {\bibinfo  {journal} {Phys. Rev. E}\
  }\textbf {\bibinfo {volume} {71}},\ \bibinfo {pages} {026111} (\bibinfo
  {year} {2005})}\BibitemShut {NoStop}%
\bibitem [{\citenamefont {Simon}\ and\ \citenamefont
  {Moustakas}(2004)}]{simon2004eigenvalue}%
  \BibitemOpen
  \bibfield  {author} {\bibinfo {author} {\bibfnamefont {S.~H.}\ \bibnamefont
  {Simon}}\ and\ \bibinfo {author} {\bibfnamefont {A.~L.}\ \bibnamefont
  {Moustakas}},\ }\href@noop {} {\bibfield  {journal} {\bibinfo  {journal}
  {Physical Review E}\ }\textbf {\bibinfo {volume} {69}},\ \bibinfo {pages}
  {065101} (\bibinfo {year} {2004})}\BibitemShut {NoStop}%
\end{thebibliography}%

\end{document}